\documentclass[]{spie}  						%>>> use for US letter paper
\usepackage[]{graphicx}
\usepackage[pdfauthor={Elodie Choquet},colorlinks]{hyperref}

\newcommand*\degr{\ensuremath{^\circ}}
\graphicspath{{/Users/choquet/Documents/Travail/Talks-Papers/2015-08-9_SPIE_SanDiego/Manuscript}}

\title{Archival Legacy Investigations of Circumstellar Environments (ALICE): Statistical assessment of point source detections}

\author{\'Elodie Choquet\supit{a} and Laurent Pueyo\supit{a} and R\'emi Soummer\supit{a} and Marshall D. Perrin\supit{a} and J. Brendan Hagan\supit{a,b} and Elena Gofas-Salas\supit{a,c}  and Abhijith Rajan\supit{d} Jonathan Aguilar\supit{e} 
\skiplinehalf
\supit{a}Space Telescope Science Institute, 3700 San Martin drive, Baltimore MD 21218, USA; \\
\supit{b} Purdue University, 610 Purdue Mall, West Lafayette, IN 47907, USA\\
\supit{c}Institut d'Optique Graduate School, 2 Avenue Augustin Fresnel, 91127 Palaiseau, France\\
\supit{d}Arizona State University, Tempe, AZ 85281, USA\\
\supit{e}Johns Hopkins University, 3400 North Charles Street, Baltimore, MD 21218, USA\\
}

\authorinfo{Contact information: \href{mailto:choquet@stsci.edu}{choquet@stsci.edu}}

\begin{document} 
  \maketitle 

%%%%%%%%%%%%%%%%%%%%%%%%%%%%%%%%%%%%%%%%%%%%%%%%%%%%%%%%%%%%% 
\begin{abstract}

The ALICE program, for Archival Legacy Investigation of Circumstellar Environment, is currently conducting a virtual survey of about 400 stars, by re-analyzing the HST-NICMOS coronagraphic archive with advanced post-processing techniques. We present here the strategy that we adopted to identify detections and potential candidates for follow-up observations, and we give a preliminary overview of our detections. We present a statistical analysis conducted to evaluate the confidence level on these detection and the completeness of our candidate search.
\end{abstract}

\keywords{HST, NICMOS, exoplanets, high-contrast imaging, post-processing techniques}

%%%%%%%%%%%%%%%%%%%%%%%%%%%%%%%%%%%%%%%%%%%%%%%%%%%%%%%%%%%%%
\section{INTRODUCTION}
\label{sec:intro}  

The development of advanced post-processing techniques based on the use of a library of instrument point spread function (PSF) images to create a synthetic PSF  that optimally subtracts the residual starlight from a target image has enable significant progress in the direct imaging of extra-solar planets over the last decade. Previous PSF subtraction techniques, mainly consisting in one-to-one image subtraction either of a reference star or of an image of the science target itself with a different orientation of the field of view, were very efficient at imaging debris disks in scattered light around young nearby stars \cite{Paresce1987,Schneider1999,Weinberger1999,Augereau1999,Kalas2004,Kalas2005a,Schneider2005,Kalas2005,Kalas2006,Schneider2006,Hines2007,Kalas2007,Kalas2007a}, but mostly failed at reaching the contrast limits needed to detect faint exoplanets \cite{Lowrance2005}. Advanced post-processing algorithms based on the linear combination of PSF images (LOCI and its variants)\cite{Lafreniere2007,Thalmann2010,Soummer2011,Pueyo2012,Marois2014} or on Principal Component Analysis (PCA)\cite{Soummer2012,Amara2012,Fergus2014} can reach deeper contrast limits with ground-based observations, with PSF diversity obtained with the Angular Differential Imaging (ADI)\cite{Marois2006} or the Spectral Differential Imaging (SDI)\cite{Marois2006a} observing strategies. They are also more efficient when using PSFs from many different stars (Reference star Differential Imaging, RDI), even when acquired with first-generation instruments on the Hubble Space Telescope (HST) and separated by large time intervals. This has been demonstrated by the re-discovery of HR~8799 planets in archival NICMOS data from 1998 by \citenum{Lafreniere2009} (b planet) and \citenum{Soummer2011} (b,c, and d planets).

These results made the community realize that a fraction of all previous results obtained with first-generation coronagraphic instruments might be out-dated by the development of these advanced post-processing techniques, and started the ALICE project (Archival Legacy Investigations of Circumstellar Environments)\footnote{HST program HST-AR-12652.01, PI R. Soummer}. The goal of this program is to consistently reprocess the NICMOS coronagraphic archive with advanced post-processing methods. NICMOS was operating on-board HST for about 8 years between 1997 and 2008, and its mid-resolution channel NIC2 (pixel size $0.076$'') was equipped with a 0.3''-radius coronagraphic mask and a Lyot stop. About 400 stars were observed during the instrument operations, mostly in the two wide-band filters F110W and F160W as part of surveys looking for debris disks and planets around nearby stars. The ALICE pipeline\cite{Choquet2014d} assembles and aligns large PSF libraries from consistent subsamples of this database (acquired with identical filters and in the same NICMOS era), that are used to process each individual targets with the KLIP algorithm\cite{Soummer2012}. This project has already revealed new images of 9 debris disks previously undetected from the NICMOS data, among which 8 had never before been imaged in scattered light (\citenum{Soummer2014}, Choquet et al. in prep.). Many point sources are also uncovered in the data. This publication reviews the detections obtained so far as part of the ALICE program, and describes the method used to statistically assess the performance of our candidate search strategy.

%\cite{Savransky2015}

%%%%%%%%%%%%%%%%%%%%%%%%%%%%%%%%%%%%%%%%%%%%%%%%%%%%%%%%%%%%%
\section{CANDIDATE SEARCH} 

\subsection{Candidate search strategy}\label{sec: search strategy}

Our candidate search strategy is based on the following considerations:
\begin{itemize}
\item High-contrast imaging surveys with ground-based instruments have revealed that giant exoplanets at large separations from their host-star are relatively rare, with occurrence rates varying between 0 and 20~\% for planets of several Jupiter masses, depending on the separation range considered, on the atmospheric models used, and on the contrast limits achieved\cite{Vigan2012,Biller2013,Chauvin2015,Meshkat2015}. In this regime, which is still un-probed by other detections techniques such as transit detection or radial velocity measurement, every new detection is critical to further constrain our planet formation models. Our objective is thus to detect all the potential companions in the archive, and adopt a conservative strategy which reports candidates even at a medium confidence level.
\item Since all the NICMOS datasets are at least 7 years old, rejection/confirmation of companion candidates by common proper motion analysis with the new generation of ground-based high-contrast imaging instruments will be straight-forward thanks to their better contrast performance at the separations probed by the ALICE program ($\sim$0.3--3''). While dedicated follow-up observations will be proposed for high-confidence candidates, having astrometric and photometric characteristics of detections at low-confidence level can be very valuable in the long term to confirm/reject candidates detected with other instruments by different teams, or be reported as false positives from contrast limits.
\item The James Webb Space Telescope (JWST) will offer coronagraphic capabilities at sensitivities and wavelength ranges unavailable from the ground from 2 to 5$\mu$m (NIRCam) and 11 to 23~$\mu$m (MIRI). Since it has a mission lifetime limited to about 5 years, the optimal use of its time for exoplanetary science would be for known companion characterization rather than candidate search and/or confirmation. It is thus critical to detect as many exoplanets as possible before JWST launch.
\end{itemize}
%MIRI: coronagraphic imaging at 10.65, 11.4, 15.5 and 23 micron

As a consequence, the companion search strategy that we adopted consists of looking for point sources in a systematic way down to small separations and low contrasts, and reporting all detections even at low signal to noise ratio (SNR). With this strategy, a fraction of the reported detections will be false positives (i.e. speckles instead of real astrophysical objects). These detections will need to be prioritized by confidence level to select only the solid candidates that will be proposed for follow-up observations. We describe in Sec.~\ref{sec:completeness detections} a statistical method to estimate the confidence level and search completeness of a given detection based on its SNR, that can be used to prioritize the ALICE candidates for further observations.

The majority of the targets in the NICMOS archive have been observed with two different orientations of the telescope (hereafter called ``roll images''), in order to subtract the star PSF by roll subtraction. When available, we make use of this characteristic to limit the fraction of false positives in our candidate search, by defining a detection as a point source identified independently in each roll image, with same astrometry within 1.5 pixels and same photometry within a factor of 3. These conservative parameters account for biases induced by the post-processing that can differ from one roll to the other (e.g. for faint point sources aligned with a diffraction spike in one roll but not in the other).

The astrometry of a detection is estimated by finding the maximum of correlation between the image and a synthetic NIMCOS PSF (generated with the Tiny TIM package\cite{Krist2011}). The raw photometry is estimated by measuring the correlation at that position between the image and the normalized synthetic PSF and by correcting it from the local background level estimated in an ring from 6 to 10 pixels radii centered on the detection. The SNR of the detection (as observed in the image, uncorrected from the processing throughput) is estimated from the raw photometry and from the local noise level per resolution element measured by the standard deviation in the same ring as above of the image convolved with a synthetic NICMOS PSF.

\subsection{Intermediate overview of the detections}\label{sec:overview det}

\begin{figure}
   \begin{center}
   \includegraphics[width=\linewidth]{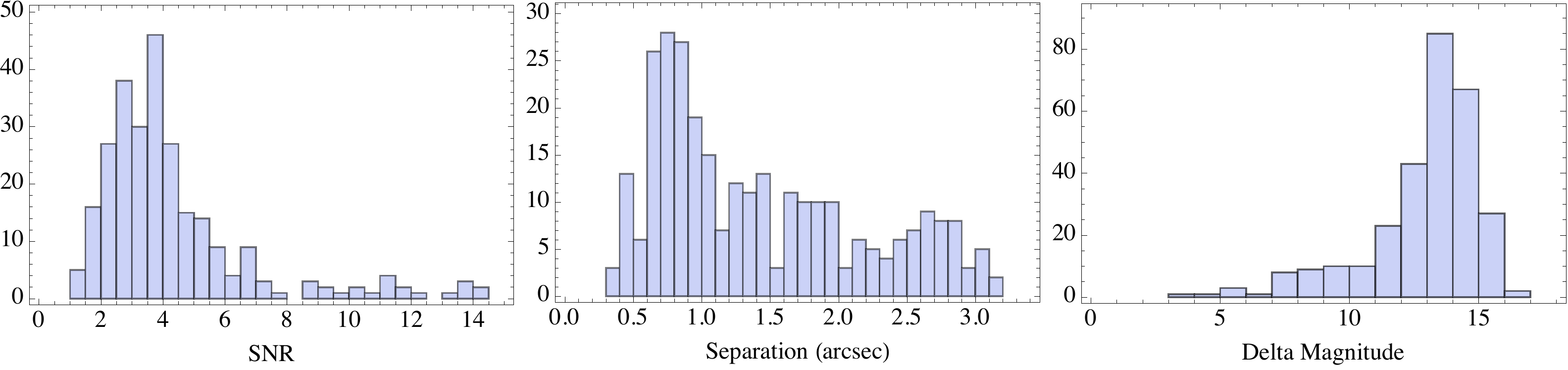}
   \end{center}
   \caption{Distributions in SNR, separation, and contrast, respectively from left to right, of the 290 detections found in the NICMOS archive to date as part of the ALICE project.They include low-contrast binaries, known background stars, and companion candidates, as well as a fraction of false positives. \label{fig:overview}}
   \end{figure}

Three large exoplanet surveys were conducted during NICMOS lifetime  (programs 7226 PI E. Becklin, 7227 PI G. Schneider, 10176 PI I. Song). To date we have re-analyzed 92\% of their targets with the final version of the ALICE pipeline. We report 237 detections in these programs, and 304 detections total in the NICMOS coronagraphic archive. Fig.~\ref{fig:overview} presents the histograms of the SNR, separation and contrast of the 290 detections for which the host star has a J or H magnitude value reported in the SIMBAD database\cite{Wenger2000}\footnote{\href{http://simbad.u-strasbg.fr/simbad}{http://simbad.u-strasbg.fr/simbad}}, which we used to compute the detection contrast from their photometry in the F110W or F160W filter using the \verb|Synphot| synthetic photometry package \footnote{Available on STSDAS website \href{http://www.stsci.edu/resources/software\_hardware/stsdas}{http://www.stsci.edu/resources/software\_hardware/stsdas}}.
 
 The detections reported here include several categories of objects: binary companions, known background stars, unknown background star / companions (real candidates), and false positive detections (i.e. speckles). About 40 objects are solid detections with SNR greater than 10, and a large fraction of the rest of this sample is detected at low SNR, short separations and high contrast, and are by-products of the candidate search strategy described in Sec.~\ref{sec: search strategy}.
 
\section{STATISTICAL ANALYSIS OF THE DETECTIONS}

In this section we describe a statistical method to estimate the efficiency of our candidate search strategy, first as a whole regardless of the detection characteristics, then specifically for candidates at a given separation and contrast. 

To optimize follow-up campaigns, our goal is to classify a detections between two categories: real astrophysical object (regardless if they are background stars, exoplanets, or binary companions) or false positives. The metric usually used for this binary classification is by setting a threshold on SNR of the detections above with it is considered as a real object instead of a speckle. We use here metrics from signal detection theory (first developed in the 1950's for radar signal detections, now extensively used in medicine to evaluate diagnostic tests) to characterize our candidate search strategy and determine an optimal SNR detection threshold. 

\begin{figure}
   \begin{center}
   \includegraphics[width=0.7\linewidth]{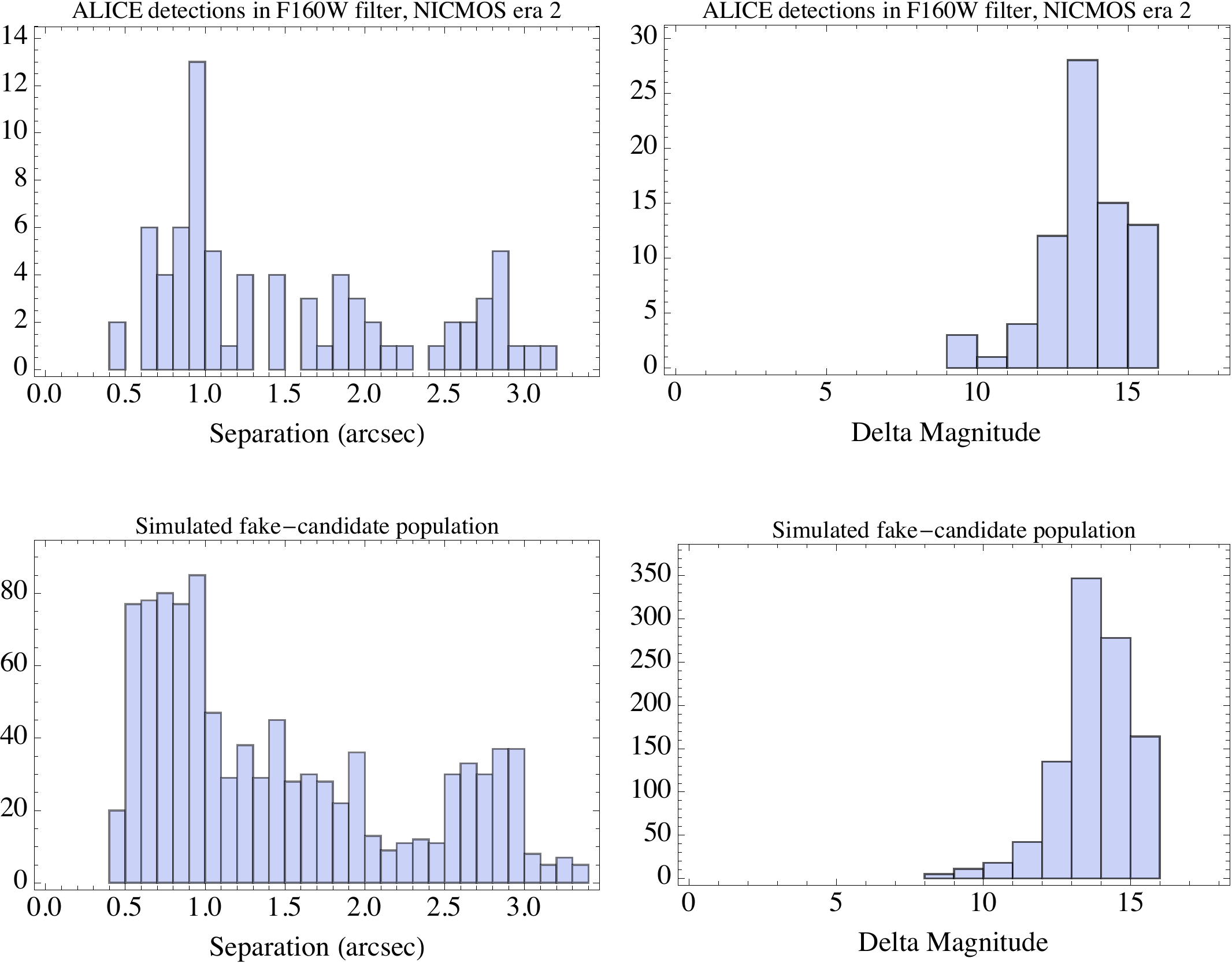}
   \end{center}
   \caption{Distributions in separation and contrast of the 76 detections found in data from the second era of NICMOS (after the replacement of its cooling system in 2002) in the F160W filter (top row), and of the 1000 fake candidates simulated to statistically assess the efficiency of our detection strategy (bottom row). The latter population has been simulated to have the same distributions in separation and contrast as the former detection list. \label{fig:fake planets}}
   \end{figure}

For this analysis, we only consider a consistent subsample of the detections reported in Sec.~\ref{sec:overview det}, restricted to datasets obtained with the F160W filter in the second era of NICMOS (after replacement of its cooling system). The histograms in separation and contrast of the detections from this subsample are presented in Fig.~\ref{fig:fake planets}, top row.

\subsection{Candidate search strategy global efficiency}

\begin{figure}
   \begin{center}
   \includegraphics[width=0.7\linewidth]{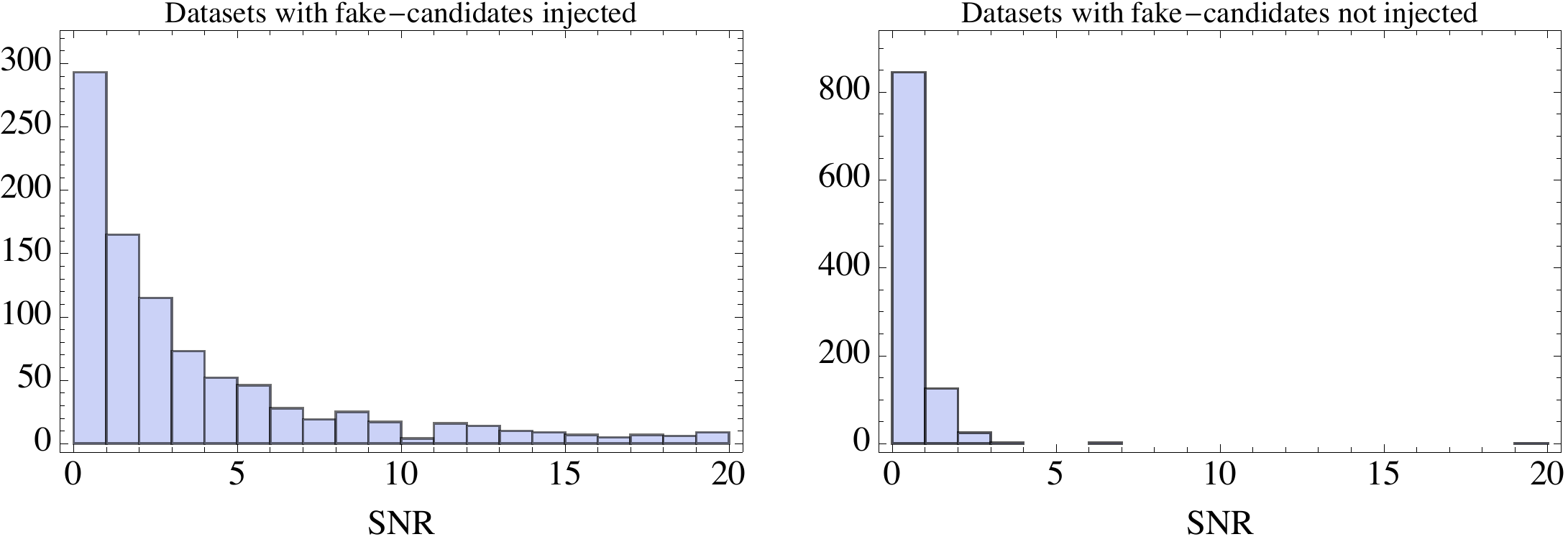}
   \end{center}
   \caption{Distributions in SNR of the fake candidates injected in NICMOS non-detection datasets (left), and of the speckle field at the same position in the datasets without the fake candidates injected (right). \label{fig:SNR fake planets}}
   \end{figure}

To evaluate our efficiency at discriminating real candidates from false positives in our sample, we perform 1000 realizations of the two following tests:
\begin{description}
\item[\textbf{Test 1}] A fake point source is injected in the NICMOS images of a target previously identified as a non-detection, at a given contrast level and sky-position. The dataset is processed with KLIP algorithm, the images are de-rotated to a common orientation with North up and combined, and the SNR of the point source is estimated, as performed with the ALICE pipeline.
\item[\textbf{Test 2}] The same non-detection images are similarly processed, de-rotated and combined, this time with no fake point source injected. The SNR of the speckle field is yet also measured at the same sky-position as in Test~1.
\end{description}
The population of fake point sources is simulated to have the same distributions in separation and contrast as the subsample of real ALICE detections presented in Fig.~\ref{sec:overview det}. The separation and contrast histograms of the fake point sources are shown in the same figure, bottom row.
Histograms  of the SNR measured for both tests are reported in Fig.~\ref{fig:SNR fake planets}. The fake point sources injected in Test 1 are detected with a wide range of SNRs, depending on their contrast and separation, while the speckles in Test 2 are all detected at low SNR.

Setting a SNR threshold above which a detection is identified as a real candidate rather than a speckle classifies the fake point sources in Test 1 as true positives and false negatives, respectively, and the speckles in Test 2 as false positives and true negatives, respectively. Fig.~\ref{fig:SNR fake planets} (left) shows the numbers of true positives and false positives detected in the 1000 realizations of these tests, as a function of the SNR threshold. 

\begin{figure}
   \begin{center}
   \includegraphics[width=0.9\linewidth]{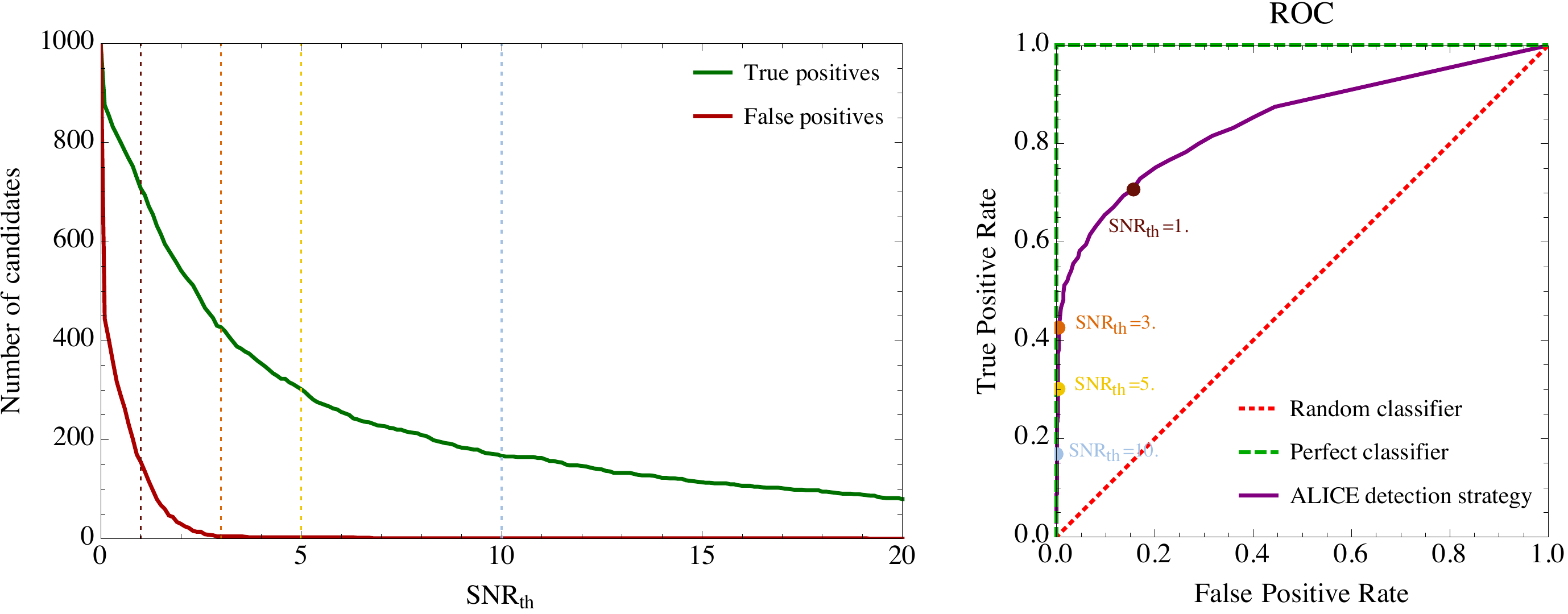}
   \end{center}
   \caption{Statistical analysis of the global efficiency of ALICE detection strategy. Left: Number of true positives (green dashed line) and of false positives (red dotted line) as a function of the SNR threshold ($SNR_{th}$) above which a detection is considered as a real candidate. Right: Receiver Operating Characteristic (ROC curve, defined by the true-positive rate as a function of false-positive rate) of ALICE candidate search strategy (purple line), compared with a completely random detection strategy (red dotted line) and a perfectly efficient detection strategy (green dashed line). Results for SNR thresholds of 1, 3, 5 and 10 are color-coded on both plots. \label{fig:ROC}}
   \end{figure}

The Receiver Operating Characteristic (ROC curve) of our candidate classification method is presented on Fig.~\ref{fig:ROC}, right. It shows the true positive rate (or sensitivity) of our method as a function of its false positive rate (or specificity) and can be used to choose the SNR threshold working point that optimizes our candidate classification. In the contrast-separation regimes probed in this analysis, setting a detection threshold at SNR$=3$ typically gives $\sim100\%$ confidence level in our detections (with few false positives) and a completeness of $\sim40\%$ (with 60\% of the real objects undetected).

The ROC presented in Fig.~\ref{fig:ROC} illustrates the performance of our systematic detection strategy, compared to a fully random detection classification (defined by equal rates of true positives and false positives). It also demonstrates the difficulty of finding all the candidates in a sample which is biased towards low-SNR point sources, generated using the distribution of our ALICE detections.

\subsection{Completeness per detection}\label{sec:completeness detections}

In this section we use the same statistical process to analyze our efficiency at detecting candidates at a given contrast and separation.  

\begin{figure}
   \begin{center}
   \includegraphics[width=0.9\linewidth]{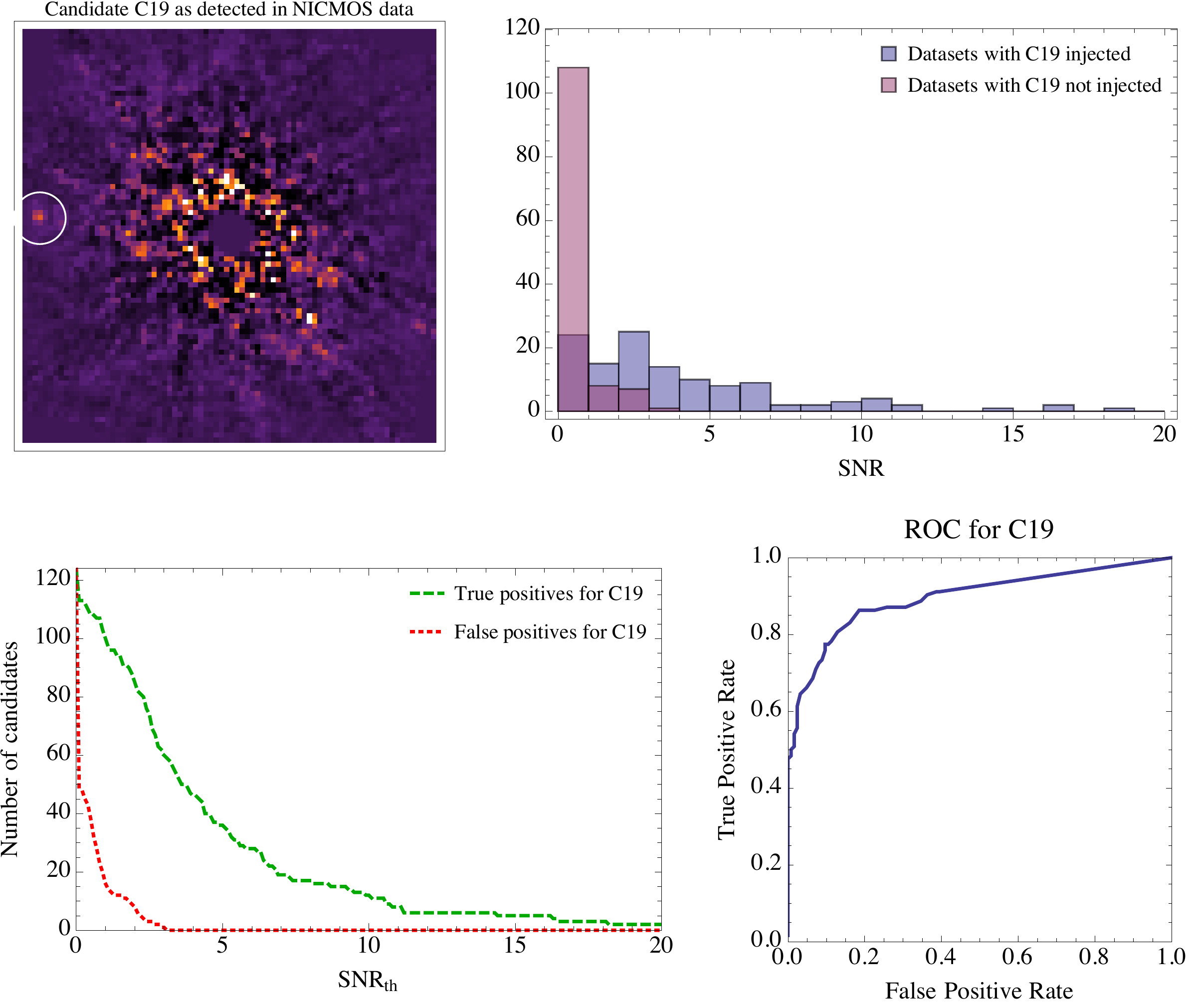}
   \end{center}
   \caption{Statistical analysis of the our efficiency at detecting a given candidate (here labeled as C19). Top-Left: Candidate C19 as seen in a NICMOS reprocessed dataset, detected at $2.8$'' from the star with a brightness 15 magnitudes fainter than the star. Top-Right: Distributions of the SNR of the detection measured when injecting C19 in 60 non-detection datasets at 2 different position angles (blue histogram), and of the speckle field measured at the same location in the datasets without injecting C19 (purple histogram). The wide range of SNR values in the first former sample reflects the difficulty to detect C19 at different position angles in a NICMOS dataset (alignment with diffraction spikes in some roll images, over-subtraction effects...). Bottom-Left: Number of true C19 positives and false C19 positives as a function of the SNR threshold above which a detection is considered as a real candidate. Bottom-Right: ROC curve for C19, showing our efficiency at detecting point sources with separations and contrasts similar to C19 in the NICMOS archive.  \label{fig:ROC C19}}
   \end{figure}

For each ALICE detection identified with the F160W filter in the second era of NICMOS (sample presented in Fig.~\ref{fig:fake planets}, top row), we realize the two following test with the 60 non-detection datasets available :
\begin{description}
\item[\textbf{Test 1}] The ALICE detection is injected in each image of the non-detection dataset, with preserved sky-position and contrast with respect to its host-star. The images are then re-processed, de-rotated and combined, and the SNR of the candidate is measured. The point source is then injected again in the same dataset at the same separation and contrast, but this time with 135$\degr$ added to its position angle, to add azimuthal diversity in the analysis (in case the point source is aligned with a diffraction spike  in some datasets), and repeat the process.
\item[\textbf{Test 2}] The non-detection dataset is similarly processed, and the SNR of the speckle field is measured at the same two positions where the ALICE candidate was injected in Test 1.
\end{description}
The true positive and false positive rates are estimated from the SNR distributions from both tests for each candidate, as a function of the SNR detection threshold. The results of these tests are presented as an example for one candidate (``C19'') in Fig.~\ref{fig:ROC C19}.

\begin{figure}
   \begin{center}
   \includegraphics[width=\linewidth]{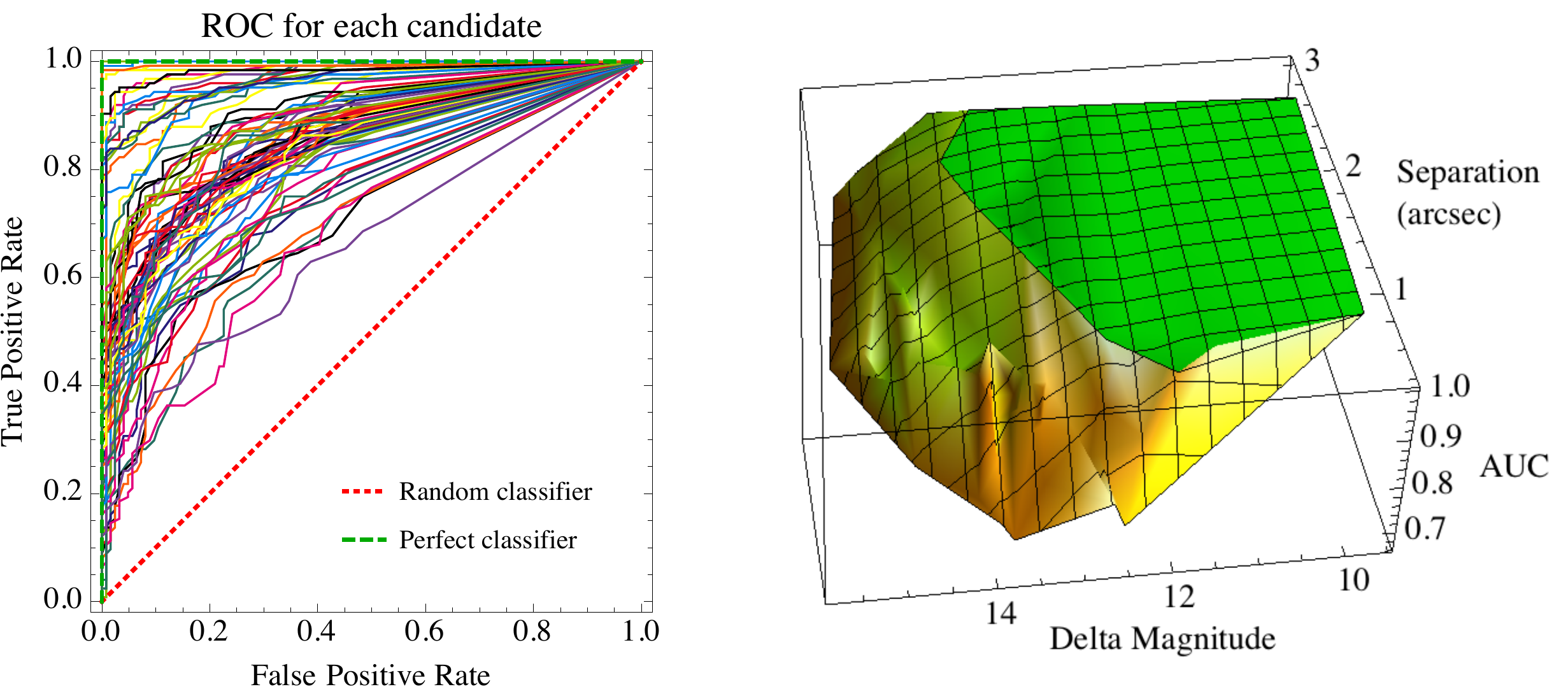}
   \end{center}
   \caption{Statistical analysis of the our efficiency at detecting each candidate. Left: ROC curve for each ALICE detection. Right: Area Under the ROC Curve (AUC) as a function of the separation and contrast of our candidates, which show our efficiency at detecting a given candidate (see text for details). The green area shows the parameter space where we are better than 95\% efficient, both in completeness and confidence level for our detections.\label{fig:AUC}}
   \end{figure}

The ROCs of all candidates are presented in Fig.~\ref{fig:AUC} (left). The candidates the most easy to detect (at large separation and low-contrast from the star) are systematically identified with Test 1, never mistaken for speckles in Test 2, and the SNR histograms from the two tests are well separated. The ROCs of these solid candidates are close to the 100\% true positive rate for 0\% false positive rate, which is characteristic of a ``perfect binary classifier''. For such candidates, the area under the ROC curve (AUC) is close to 1, the maximum value. On the other hand, candidates at small separation and high contrast are easily confused with speckles and present the greatest challenge. Their SNR histograms from Test 1 and Test 2 are largely overlapping, and their ROC curves are close to the equal rate of false positive and true positive detections -- characteristic of a ``random classifier'' with an AUC value of 0.5.

For a point source at given contrast and separation, the AUC value provides a metric that evaluates the combination of our completeness (ability to detect all the true candidates in the sample) and our confidence level (tendency to include false positives in our sample) in our detections. Fig.~\ref{fig:AUC} (right) presents the AUC values of the ALICE detections as a function of their separations and contrasts from their host-star. The green area shows the part of this parameter space where we are the most efficient in our detection strategy, with $\mathrm{AUC} > 95\% $. At large separation ($\sim3$''), we are complete in this regime down to contrast of delta magnitude $\sim14$. At low contrast ($\Delta m \sim10$), we are complete down to a separation of $\sim0.8$''.% Beyond this part of the parameter space, the probability that our detections include false alarms or miss true candidates would be non-negligible if no selection on the detection SNRs is applied.

\section{Conclusion}

We presented a preliminary overview of the detections found as part of the ALICE project by re-analyzing the NICMOS coronagraphic archive. We described the candidate search strategy that led to these detections, which include point sources at low SNR, short separations and high-contrast. To prioritize these detections for future follow-up observations, we conducted a statistical analysis consisting in injecting fake point sources with specific characteristics in non-detection NICMOS datasets to estimate true positive and false positive rates from these simulations. We found with this method that our detection strategy is complete with a high confidence level for separations down to $\sim0.8$'' at low-contrast levels, and for delta magnitude down to $\sim14$ at 3'' separation from the star.

%%%%%%%%%%%%%%%%%%%%%%%%%%%%%%%%%%%%%%%%%%%%%%%%%%%%%%%%%%%%%
\acknowledgments     %>>>> equivalent to \section*{ACKNOWLEDGMENTS}       
This project was made possible by the Mikulski Archive for Space Telescopes (MAST) at STScI. Support was provided by NASA through grants HST-AR-12652.01 (PI R. Soummer), HST-GO-11136.09-A (PI D. Golimowski), HST-GO-13855 (PI E. Choquet), and by STScI Director's Discretionary Research funds, from STScI, which is operated by AURA under NASA contract NAS5-26555. 
This research has made use of the SIMBAD database, operated at CDS, Strasbourg, France, and of the Synphot package distributed as part of STSDAS which is a product of STScI.
%%%%%%%%%%%%%%%%%%%%%%%%%%%%%%%%%%%%%%%%%%%%%%%%%%%%%%%%%%%%%
%%%%% References %%%%%

\bibliography{biblio-disk-planet.bib}   %>>>> bibliography data in report.bib

\begin{thebibliography}{10}

\bibitem{Paresce1987}
{Paresce}, F. and {Burrows}, C., ``{Broad-band imaging of the Beta Pictoris
  circumstellar disk},'' {\em \apjl}~{\bf 319},  L23--L25 (Aug. 1987).

\bibitem{Schneider1999}
{Schneider}, G., {Smith}, B.~A., {Becklin}, E.~E., {Koerner}, D.~W., {Meier},
  R., {Hines}, D.~C., {Lowrance}, P.~J., {Terrile}, R.~J., {Thompson}, R.~I.,
  and {Rieke}, M., ``{NICMOS Imaging of the HR 4796A Circumstellar Disk},''
  {\em \apjl}~{\bf 513},  L127--L130 (Mar. 1999).

\bibitem{Weinberger1999}
{Weinberger}, A.~J., {Becklin}, E.~E., {Schneider}, G., {Smith}, B.~A.,
  {Lowrance}, P.~J., {Silverstone}, M.~D., {Zuckerman}, B., and {Terrile},
  R.~J., ``{The Circumstellar Disk of HD 141569 Imaged with NICMOS},'' {\em
  \apjl}~{\bf 525},  L53--L56 (Nov. 1999).

\bibitem{Augereau1999}
{Augereau}, J.~C., {Lagrange}, A.~M., {Mouillet}, D., and {M{\'e}nard}, F.,
  ``{HST/NICMOS2 observations of the HD 141569 A circumstellar disk},'' {\em
  \aap}~{\bf 350},  L51--L54 (Oct. 1999).

\bibitem{Kalas2004}
{Kalas}, P., {Liu}, M.~C., and {Matthews}, B.~C., ``{Discovery of a Large Dust
  Disk Around the Nearby Star AU Microscopii},'' {\em Science}~{\bf 303},
  1990--1992 (Mar. 2004).

\bibitem{Kalas2005a}
{Kalas}, P., {Graham}, J.~R., and {Clampin}, M., ``{A planetary system as the
  origin of structure in Fomalhaut's dust belt},'' {\em \nat}~{\bf 435},
  1067--1070 (June 2005).

\bibitem{Schneider2005}
{Schneider}, G., {Silverstone}, M.~D., and {Hines}, D.~C., ``{Discovery of a
  Nearly Edge-on Disk around HD 32297},'' {\em \apjl}~{\bf 629},  L117--L120
  (Aug. 2005).

\bibitem{Kalas2005}
{Kalas}, P., ``{First Optical Images of Circumstellar Dust Surrounding the
  Debris Disk Candidate HD 32297},'' {\em \apjl}~{\bf 635},  L169--L172 (Dec.
  2005).

\bibitem{Kalas2006}
{Kalas}, P., {Graham}, J.~R., {Clampin}, M.~C., and {Fitzgerald}, M.~P.,
  ``{First Scattered Light Images of Debris Disks around HD 53143 and HD
  139664},'' {\em \apjl}~{\bf 637},  L57--L60 (Jan. 2006).

\bibitem{Schneider2006}
{Schneider}, G., {Silverstone}, M.~D., {Hines}, D.~C., {Augereau}, J.-C.,
  {Pinte}, C., {M{\'e}nard}, F., {Krist}, J., {Clampin}, M., {Grady}, C.,
  {Golimowski}, D., {Ardila}, D., {Henning}, T., {Wolf}, S., and {Rodmann}, J.,
  ``{Discovery of an 86 AU Radius Debris Ring around HD 181327},'' {\em
  \apj}~{\bf 650},  414--431 (Oct. 2006).

\bibitem{Hines2007}
{Hines}, D.~C., {Schneider}, G., {Hollenbach}, D., {Mamajek}, E.~E.,
  {Hillenbrand}, L.~A., {Metchev}, S.~A., {Meyer}, M.~R., {Carpenter}, J.~M.,
  {Moro-Mart{\'{\i}}n}, A., {Silverstone}, M.~D., {Kim}, J.~S., {Henning}, T.,
  {Bouwman}, J., and {Wolf}, S., ``{The Moth: An Unusual Circumstellar
  Structure Associated with HD 61005},'' {\em \apjl}~{\bf 671},  L165--L168
  (Dec. 2007).

\bibitem{Kalas2007}
{Kalas}, P., {Duchene}, G., {Fitzgerald}, M.~P., and {Graham}, J.~R.,
  ``{Discovery of an Extended Debris Disk around the F2 V Star HD 15745},''
  {\em \apjl}~{\bf 671},  L161--L164 (Dec. 2007).

\bibitem{Kalas2007a}
{Kalas}, P., {Fitzgerald}, M.~P., and {Graham}, J.~R., ``{Discovery of Extreme
  Asymmetry in the Debris Disk Surrounding HD 15115},'' {\em \apjl}~{\bf 661},
  L85--L88 (May 2007).

\bibitem{Lowrance2005}
{Lowrance}, P.~J., {Becklin}, E.~E., {Schneider}, G., {Kirkpatrick}, J.~D.,
  {Weinberger}, A.~J., {Zuckerman}, B., {Dumas}, C., {Beuzit}, J.-L., {Plait},
  P., {Malumuth}, E., {Heap}, S., {Terrile}, R.~J., and {Hines}, D.~C., ``{An
  Infrared Coronagraphic Survey for Substellar Companions},'' {\em \aj}~{\bf
  130},  1845--1861 (Oct. 2005).

\bibitem{Lafreniere2007}
{Lafreni{\`e}re}, D., {Marois}, C., {Doyon}, R., {Nadeau}, D., and {Artigau},
  {\'E}., ``{A New Algorithm for Point-Spread Function Subtraction in
  High-Contrast Imaging: A Demonstration with Angular Differential Imaging},''
  {\em \apj}~{\bf 660},  770--780 (May 2007).

\bibitem{Thalmann2010}
{Thalmann}, C., {Grady}, C.~A., {Goto}, M., {Wisniewski}, J.~P., {Janson}, M.,
  {Henning}, T., {Fukagawa}, M., {Honda}, M., {Mulders}, G.~D., {Min}, M.,
  {Moro-Mart{\'{\i}}n}, A., {McElwain}, M.~W., {Hodapp}, K.~W., {Carson}, J.,
  {Abe}, L., {Brandner}, W., {Egner}, S., {Feldt}, M., {Fukue}, T., {Golota},
  T., {Guyon}, O., {Hashimoto}, J., {Hayano}, Y., {Hayashi}, M., {Hayashi}, S.,
  {Ishii}, M., {Kandori}, R., {Knapp}, G.~R., {Kudo}, T., {Kusakabe}, N.,
  {Kuzuhara}, M., {Matsuo}, T., {Miyama}, S., {Morino}, J.-I., {Nishimura}, T.,
  {Pyo}, T.-S., {Serabyn}, E., {Shibai}, H., {Suto}, H., {Suzuki}, R.,
  {Takami}, M., {Takato}, N., {Terada}, H., {Tomono}, D., {Turner}, E.~L.,
  {Watanabe}, M., {Yamada}, T., {Takami}, H., {Usuda}, T., and {Tamura}, M.,
  ``{Imaging of a Transitional Disk Gap in Reflected Light: Indications of
  Planet Formation Around the Young Solar Analog LkCa 15},'' {\em \apjl}~{\bf
  718},  L87--L91 (Aug. 2010).

\bibitem{Soummer2011}
{Soummer}, R., {Brendan Hagan}, J., {Pueyo}, L., {Thormann}, A., {Rajan}, A.,
  and {Marois}, C., ``{Orbital Motion of HR 8799 b, c, d Using Hubble Space
  Telescope Data from 1998: Constraints on Inclination, Eccentricity, and
  Stability},'' {\em \apj}~{\bf 741},  55 (Nov. 2011).

\bibitem{Pueyo2012}
{Pueyo}, L., {Crepp}, J.~R., {Vasisht}, G., {Brenner}, D., {Oppenheimer},
  B.~R., {Zimmerman}, N., {Hinkley}, S., {Parry}, I., {Beichman}, C.,
  {Hillenbrand}, L., {Roberts}, L.~C., {Dekany}, R., {Shao}, M., {Burruss}, R.,
  {Bouchez}, A., {Roberts}, J., and {Soummer}, R., ``{Application of a Damped
  Locally Optimized Combination of Images Method to the Spectral
  Characterization of Faint Companions Using an Integral Field Spectrograph},''
  {\em \apjs}~{\bf 199},  6 (Mar. 2012).

\bibitem{Marois2014}
{Marois}, C., {Correia}, C., {Galicher}, R., {Ingraham}, P., {Macintosh}, B.,
  {Currie}, T., and {De Rosa}, R., ``{GPI PSF subtraction with TLOCI: the next
  evolution in exoplanet/disk high-contrast imaging},'' in [{\em
  \procspie}{\nolinebreak\hspace{0.1em}]},  {\em \procspie} {\bf 9148},  0
  (July 2014).

\bibitem{Soummer2012}
{Soummer}, R., {Pueyo}, L., and {Larkin}, J., ``{Detection and Characterization
  of Exoplanets and Disks Using Projections on Karhunen-Lo{\`e}ve
  Eigenimages},'' {\em \apjl}~{\bf 755},  L28 (Aug. 2012).

\bibitem{Amara2012}
{Amara}, A. and {Quanz}, S.~P., ``{PYNPOINT: an image processing package for
  finding exoplanets},'' {\em \mnras}~{\bf 427},  948--955 (Dec. 2012).

\bibitem{Fergus2014}
{Fergus}, R., {Hogg}, D.~W., {Oppenheimer}, R., {Brenner}, D., and {Pueyo}, L.,
  ``{S4: A Spatial-spectral model for Speckle Suppression},'' {\em \apj}~{\bf
  794},  161 (Oct. 2014).

\bibitem{Marois2006}
{Marois}, C., {Lafreni{\`e}re}, D., {Doyon}, R., {Macintosh}, B., and {Nadeau},
  D., ``{Angular Differential Imaging: A Powerful High-Contrast Imaging
  Technique},'' {\em \apj}~{\bf 641},  556--564 (Apr. 2006).

\bibitem{Marois2006a}
{Marois}, C., {Phillion}, D.~W., and {Macintosh}, B., ``{Exoplanet detection
  with simultaneous spectral differential imaging: effects of
  out-of-pupil-plane optical aberrations},'' in [{\em
  \procspie}{\nolinebreak\hspace{0.1em}]},  {\em \procspie} {\bf 6269} (July
  2006).

\bibitem{Lafreniere2009}
{Lafreni{\`e}re}, D., {Marois}, C., {Doyon}, R., and {Barman}, T.,
  ``{HST/NICMOS Detection of HR 8799 b in 1998},'' {\em \apjl}~{\bf 694},
  L148--L152 (Apr. 2009).

\bibitem{Choquet2014d}
{Choquet}, {\'E}., {Pueyo}, L., {Hagan}, J.~B., {Gofas-Salas}, E., {Rajan}, A.,
  {Chen}, C., {Perrin}, M.~D., {Debes}, J., {Golimowski}, D., {Hines}, D.~C.,
  {N'Diaye}, M., {Schneider}, G., {Mawet}, D., {Marois}, C., and {Soummer}, R.,
  ``{Archival legacy investigations of circumstellar environments: overview and
  first results},'' in [{\em \procspie}{\nolinebreak\hspace{0.1em}]},  {\em
  \procspie} {\bf 9143},  57 (Aug. 2014).

\bibitem{Soummer2014}
{Soummer}, R., {Perrin}, M.~D., {Pueyo}, L., {Choquet}, {\'E}., {Chen}, C.,
  {Golimowski}, D.~A., {Brendan Hagan}, J., {Mittal}, T., {Moerchen}, M.,
  {N'Diaye}, M., {Rajan}, A., {Wolff}, S., {Debes}, J., {Hines}, D.~C., and
  {Schneider}, G., ``{Five Debris Disks Newly Revealed in Scattered Light from
  the Hubble Space Telescope NICMOS Archive},'' {\em \apjl}~{\bf 786},  L23
  (May 2014).

\bibitem{Vigan2012}
{Vigan}, A., {Patience}, J., {Marois}, C., {Bonavita}, M., {De Rosa}, R.~J.,
  {Macintosh}, B., {Song}, I., {Doyon}, R., {Zuckerman}, B., {Lafreni{\`e}re},
  D., and {Barman}, T., ``{The International Deep Planet Survey. I. The
  frequency of wide-orbit massive planets around A-stars},'' {\em \aap}~{\bf
  544},  A9 (Aug. 2012).

\bibitem{Biller2013}
{Biller}, B.~A., {Liu}, M.~C., {Wahhaj}, Z., {Nielsen}, E.~L., {Hayward},
  T.~L., {Males}, J.~R., {Skemer}, A., {Close}, L.~M., {Chun}, M., {Ftaclas},
  C., {Clarke}, F., {Thatte}, N., {Shkolnik}, E.~L., {Reid}, I.~N., {Hartung},
  M., {Boss}, A., {Lin}, D., {Alencar}, S.~H.~P., {de Gouveia Dal Pino}, E.,
  {Gregorio-Hetem}, J., and {Toomey}, D., ``{The Gemini/NICI Planet-Finding
  Campaign: The Frequency of Planets around Young Moving Group Stars},'' {\em
  \apj}~{\bf 777},  160 (Nov. 2013).

\bibitem{Chauvin2015}
{Chauvin}, G., {Vigan}, A., {Bonnefoy}, M., {Desidera}, S., {Bonavita}, M.,
  {Mesa}, D., {Boccaletti}, A., {Buenzli}, E., {Carson}, J., {Delorme}, P.,
  {Hagelberg}, J., {Montagnier}, G., {Mordasini}, C., {Quanz}, S.~P.,
  {Segransan}, D., {Thalmann}, C., {Beuzit}, J.-L., {Biller}, B., {Covino}, E.,
  {Feldt}, M., {Girard}, J., {Gratton}, R., {Henning}, T., {Kasper}, M.,
  {Lagrange}, A.-M., {Messina}, S., {Meyer}, M., {Mouillet}, D., {Moutou}, C.,
  {Reggiani}, M., {Schlieder}, J.~E., and {Zurlo}, A., ``{The VLT/NaCo large
  program to probe the occurrence of exoplanets and brown dwarfs at wide
  orbits. II. Survey description, results, and performances},'' {\em \aap}~{\bf
  573},  A127 (Jan. 2015).

\bibitem{Meshkat2015}
{Meshkat}, T., {Kenworthy}, M.~A., {Reggiani}, M., {Quanz}, S.~P., {Mamajek},
  E.~E., and {Meyer}, M.~R., ``{Searching for gas giant planets on Solar System
  scales - A NACO/APP L'-band survey of A- and F-type Main Sequence stars},''
  {\em ArXiv e-prints}  (Aug. 2015).

\bibitem{Krist2011}
{Krist}, J.~E., {Hook}, R.~N., and {Stoehr}, F., ``{20 years of Hubble Space
  Telescope optical modeling using Tiny Tim},'' in [{\em
  \procspie}{\nolinebreak\hspace{0.1em}]},  {\em \procspie} {\bf 8127} (Sept.
  2011).

\bibitem{Wenger2000}
{Wenger}, M., {Ochsenbein}, F., {Egret}, D., {Dubois}, P., {Bonnarel}, F.,
  {Borde}, S., {Genova}, F., {Jasniewicz}, G., {Lalo{\"e}}, S., {Lesteven}, S.,
  and {Monier}, R., ``{The SIMBAD astronomical database. The CDS reference
  database for astronomical objects},'' {\em \aaps}~{\bf 143},  9--22 (Apr.
  2000).

\end{thebibliography}
\bibliographystyle{spiebib}   %>>>> makes bibtex use spiebib.bst

\end{document}